\documentclass[aip,preprint,aps,amsmath,amssymb]{revtex4-2}

\usepackage{graphicx}
\usepackage{amsmath,amssymb,bm,epsfig}
\usepackage{dcolumn}
\usepackage{bm}
\usepackage{xcolor}

\begin{document}

\preprint{APL Quantum 1, 046120 (2024)}

\title{Variational approach to atom-membrane dynamics}

\author{Dennis P. Clougherty}
\email{dennis.clougherty@uvm.edu}

\affiliation{
Department of Physics, 
University of Vermont\\
Burlington, VT 05405-0125}

\date{November 6, 2024}

\begin{abstract}
Using the Dirac-Frenkel variational principle, a time-dependent description of the dynamics of a two-level system coupled to a bosonic bath is formulated.  The method is applied to the case of a gas of cold atoms adsorbing to an elastic membrane at finite temperature via phonon creation.  The time-dependence of the system state is analytically calculated using Laplace transform methods, and a closed-form expression for the transition rate is obtained.  Atoms in the gas transition to the adsorbed state through a  resonance that has contributions from  a  distribution of vibrational modes of the membrane.  The resonance can decay with the creation of a phonon to complete the adsorption process.  The adsorption rate  at low membrane temperatures agrees with the golden rule estimate to lowest order in the coupling constant for values greater than a critical coupling strength.   Below this critical coupling strength, the adsorption rate is exponentially suppressed by a phonon reduction factor whose exponent diverges with increasing adsorbent size.   The rate changes discontinuously with coupling strength for low temperature membranes, and the magnitude of the discontinuity decreases with increasing temperature.  These variational results suggest the quantum adsorption model may contain a first-order quantum phase transition. 
 \end{abstract}

\maketitle
 
\newpage

\section{Introduction}

The surface-catalyzed reaction is a workhorse of science; it provides a way to accelerate reaction rates and selectively produce commercially desired reaction products in quantity.  In the emerging field of cryochemistry, conducting chemical reactions at ultracold temperatures offers the possibility of controlling the quantum state of the reaction product; for example, it is now possible to experimentally prepare and manipulate the quantum state of diatomic molecules of alkali atoms \cite{ye} from a mixture of monatomic gases.  It is the hope of researchers that such control will be used in a variety of emerging quantum technologies \cite{superchem}.  To expedite such quantum-controlled reactions, ultracold quantum catalysis on surfaces may well play a role.        

To assess the prospects for quantum catalysis, nonperturbative studies of ultracold atomic and molecular scattering and adsorption can provide foundational information.    The adsorption of ultracold atoms differs from the adsorption of atoms at room temperature.  It has been established both theoretically \cite{dpc92} and experimentally \cite{yu93} that the adsorption rate of cold atoms can be dramatically suppressed in comparison to rates at elevated temperatures by quantum mechanical effects: (1) quantum reflection \cite{dpc92, cole-review} of the cold atoms away from the surface, and (2) a phonon orthogonality catastrophe \cite{dpc11} resulting from the surface displacement that accompanies adsorption.  The first effect is a single-particle phenomenon that depends on the wave mechanics of the cold atoms; the second effect is a many-body phenomenon that results from the behavior of the phonon matrix element between the initial and final states of the surface.  In the most extreme case of physisorption on a 2D material, it has been proposed \cite{dpc13} that this phonon reduction factor can completely suppress cold atom adsorption on suspended graphene.  

In this work, a time-dependent, nonperturbative description of phonon-assisted cold atom adsorption is presented.  Using the Dirac-Frenkel variational principle  \cite{dirac, frenkel}, a closed-form expression for adsorption rate can be obtained.  One strength of this method is that the case of intermediate atom-phonon coupling can be treated where the adsorption rate changes discontinuously with atom-phonon coupling strength.  The framework presented can be customized in a straightforward way to describe a variety of reactions in the quantum regime.

\section{Quantum adsorption model}

The following Hamiltonian, derived in Ref.~\cite{dpc13}, gives the basic model for describing phonon-assisted quantum adsorption
\begin{equation}
H=H_a+H_p+H_{i1}+H_{i2}
\end{equation}
where
\begin{eqnarray}
H_a&=& E_c c^\dagger c -E_b b^\dagger b\\
H_p&=& \sum_n \omega_n a^\dagger_n a_n\\
H_{i1}&=&-g_{kb}(c^\dagger b+b^\dagger c) \sum_n  (a^\dagger_n +a_n)\\
H_{i2}&=&-g_{bb} b^\dagger b \sum_n  (a^\dagger_n +a_n)
\label{model}
\end{eqnarray} 


The model treats two states of the atom: the first is the initial state of a cold atom in the gas phase with  energy $E_c$; the second is the atom bound to the surface with energy $-E_b$.  $c^\dagger$ ($c$) creates (annihilates) an atom in the gas phase, while $b^\dagger$ ($b$) creates (annihilates) an atom bound to the surface. $a_n^\dagger$ ($a_n$) creates (annihilates) a phonon from the nth mode. Atom adsorption occurs by  displacement of the elastic membrane which is assumed to be under tension.  

The coupling constants $g_{kb}$ and $g_{bb}$ are atom matrix elements of the normal derivative of the static surface potential, as described in Ref. \cite{dpc13}. For cold atoms, $g_{kb}$ varies as the square root of the atom energy in the gas phase \cite{sengupta} and can be easily tuned experimentally \cite{yu93}. The membrane is taken to be a disk of radius $a$ that is clamped at its edge.   

For the case where $g_{bb}\ll g_{kb}$, the dynamics is expected to be contained in the multimode Rabi model \cite{mqrm,rabi-model}
\begin{equation}
H_R=H_a+H_p+H_{i1}
\end{equation} 

The transition rate to the adsorbed state via single phonon emission is given by the golden rule for weak coupling $g_{kb}\ll E_b$ 
\begin{equation}
R_0=2\pi \rho g_{kb}^2(n(\omega_s)+1)
\label{gr}
\end{equation}
where $\omega_s=E_c+E_b$, $n(E)$ is the thermally-averaged number of phonons with energy $E$ present in the membrane in equilibrium at temperature $T$ and $\rho$ is the (constant) vibrational density of states for circularly symmetric modes.  


For the case where $g_{bb}\gg g_{kb}$, the dynamics should be described by the independent boson model \cite{mahan} $H_{\rm IBM}$  subject to the perturbation of $H_{i1}$.  
\begin{equation}
H_{\rm IBM}=H_a+H_p+H_{i2}
\end{equation}
In this regime, the transition rate behaves \cite{dpc-IBM} as
\begin{equation}
R_{\rm IBM}\approx R_0 \exp(-2 F)
\end{equation}
where $F=\frac{1}{2}\sum_p \big({g_{bb}\over \omega_p}\big)^2 \coth({\beta\omega_p\over 2})$.  The Franck-Condon-like factor $\exp(-2 F)$ is a result of the displacement of the membrane in the presence of the bound atom.  For large membranes, this phonon reduction factor behaves \cite{dpc-IBM} as $\exp(-2 F)\sim \exp(-\frac{g_{bb}^2 \rho}{\epsilon})$ where $\epsilon$ is lowest vibrational frequency supported by the membrane. 
Thus in this regime, the transition rate vanishes in the thermodynamic limit where the membrane is of infinite size ($a\to \infty$, $\epsilon\to 0$).

\section{Dirac-Frenkel variational method}

To obtain the transition rate over a range of the coupling constants that bridges these extreme cases, a non-perturbative approach is required.  In this work, the time-dependent variational method of Dirac and Frenkel \cite{dirac, frenkel} is applied.  A time-dependent variational state that describes the fundamental adsorption process is chosen, and an effective Lagrangian is obtained for the system.   Time-dependent amplitudes in the variational state serve as generalized coordinates. Equations of motion for the variational amplitudes follow from the Euler-Lagrange equations for the effective Lagrangian.   The equations of motion are subsequently solved using integral transform methods.

The variational state of the system is taken to be a superposition of two types of states:  the initial state of the atom of energy $E_c$ with a thermal distribution of phonons supported on the membrane; and secondly, an  surface polaron consisting of the  atom bound to the displaced membrane with a phonon present.

\begin{equation}
|\psi(t)\rangle=\bigg(C(t) c^\dagger+\sum_m B_m(t) {\cal D}(\{ \lambda\}) a_m^\dagger b^\dagger\bigg) |{\rm \{n_q\}}\rangle
\label{psi}
\end{equation}
where the membrane displacement operator is ${\cal D}(\{ \lambda\})=\exp\big(\sum_p \lambda_p(a_p^\dagger-a_p)  \big)$.   The set of $\{\lambda_p\}_{p=1}^N$ are variational parameters that specify the modal displacements.  The phonon states are defined by the set of occupation numbers $\{n_q\}$ such that $|{\rm \{n_q\}}\rangle=\prod_q \frac{(a_q^\dagger)^{n_q}}{\sqrt{n_q!}}|0\rangle$. $C(t)$ and the set of $B_n(t)$ is taken to be variational functions.

The effective Lagrangian in the Dirac-Frenkel approach is given by
\begin{equation}
L=\langle\psi(t)|(i{d\over dt} - H)|\psi(t)\rangle
\label{DF}
\end{equation}
{Phonon disentanglement of the resulting matrix elements from substitution of Eq.~\ref{psi} into Eq.~\ref{DF} can be facilitated with the Baker-Campbell-Hausdorff formula and the use of commutation relations for the displacement operator, e.g., $[a_m, {\cal D}(\{ \lambda\})]=\lambda_m {\cal D}(\{ \lambda\})$. (See Appendix of Ref.~\cite{dpc-IBM} for useful identities in evaluating thermal averages.)  }
The following Lagrangian results 
\begin{eqnarray}
L&=&iC^*{dC\over dt}+i\sum_p (n_p+1)B_p^*{dB_p\over dt}-(E_c+\sum_m n_m\omega_m) C^* C\nonumber\\
&+&
\sum_m B_m^* B_m(n_m+1)(E_b+\Delta-\omega_m-\sum_p n_p\omega_p)\nonumber\\
&+&{\bar g}\sum_m(C^*B_m+B_m^*C)(n_m+1-\lambda_m\sum_p \lambda_p(n_p+1))
\label{lagrangian}
\end{eqnarray}
where the polaron shift $\Delta\equiv g_{bb}\sum_p (2\lambda_p-\lambda_p^2\omega_p/g_{bb})$ {and the effective transitional coupling constant ${\bar g}=g_{kb}\exp(-\frac{1}{2}\sum_p\lambda_p^2(2 n_p+1))$}.  

\section{Euler-Lagrange equations}
The {dynamical equations for the variational functions are the} Euler-Lagrange equations derived  from {Eq.~\ref{lagrangian}}.

\begin{eqnarray}
\label{euler-lagrange1}
i{dC\over dt}&=&(E_c+\sum_p n_p\omega_p)C-{\bar g}\sum_m (n_m+1-\lambda_m\sum_p \lambda_p(n_p+1))B_m\\
i{dB_n\over dt}&=&-(E_b+\Delta-\omega_n-\sum_p n_p\omega_p)B_n-{\bar g} {(n_n+1-\lambda_n\sum_p \lambda_p(n_p+1))\over(n_n+1)}C
\label{euler-lagrange2}
\end{eqnarray}

The membrane displacement generated by  $\{\lambda_p\}$ are fixed by minimizing the free energy of the system.
The following equation for the modal displacements results
\begin{equation}
\lambda_p^{-1}=\frac{\omega_p}{g_{bb}}+\alpha\coth({\beta\omega_p\over 2})\big(\sum_n \lambda_n\big)^2\exp\big(-\sum_n {\lambda_n^2}\coth({\beta\omega_n\over 2})\big)
\label{lambda}
\end{equation}
where $\alpha={g_{kb}^2\over g_{bb}E_b}$.   
The dynamics is found in the solution of this set of coupled linear first-order equations for the variational functions subject to the initial conditions ($C(0)=1$ and $B_n(0)=0$).  

A self-consistent solution to Eq.~\ref{lambda} can efficiently be obtained by iteration.  A field $\Delta_0$ is defined
\begin{equation}
\Delta_0=\big(\sum_n \lambda_n\big)^2\exp\big(-\sum_n {\lambda_n^2}\coth({\beta\omega_n\over 2})\big)
\label{Delta}
\end{equation}
The modal displacements can then expressed in terms of $\Delta_0$
\begin{equation}
\lambda_p={1\over \frac{\omega_p}{g_{bb}}+\alpha\Delta_0 \coth({\beta\omega_p\over 2})}
\label{sclambda}
\end{equation}
Eqs.~\ref{Delta} and \ref{sclambda} are then iterated until self-consistency is achieved.

Eqs.~\ref{euler-lagrange1} and \ref{euler-lagrange2} can be solved analytically with Laplace transforms; for example, the amplitude for the entrance channel $C(t)$ is   
\begin{equation}
{\tilde C(s)}={{i}\over {i s-E_c -\sum_p n_p\omega_p-{\bar g}^2\sum_p{(n_p+1-\lambda_p\sum_m \lambda_m(n_m+1))^2\over (n_p+1)(i s+E_b+\Delta-\omega_p-\sum_m n_m\omega_m)}}}
\label{C(s)}
\end{equation}
 The time-dependent amplitudes can be obtained by inverse transforming using the Bromwich contour in the complex s-plane.  
The poles of ${\tilde C(s)}$ are $N+1$ solutions to 
\begin{equation}
i s_n-E_c -\sum_p n_p\omega_p-{\bar g}^2 \Sigma(i s_n)=0
\end{equation}
where 
\begin{equation}
\Sigma(i s)=\sum_p{(n_p+1-\lambda_p\sum_m \lambda_m(n_m+1))^2\over (n_p+1)(i s+E_b+\Delta-\omega_p-\sum_m n_m\omega_m)}
\label{selfenergy}
\end{equation}

\begin{figure}
\includegraphics[width=12cm]{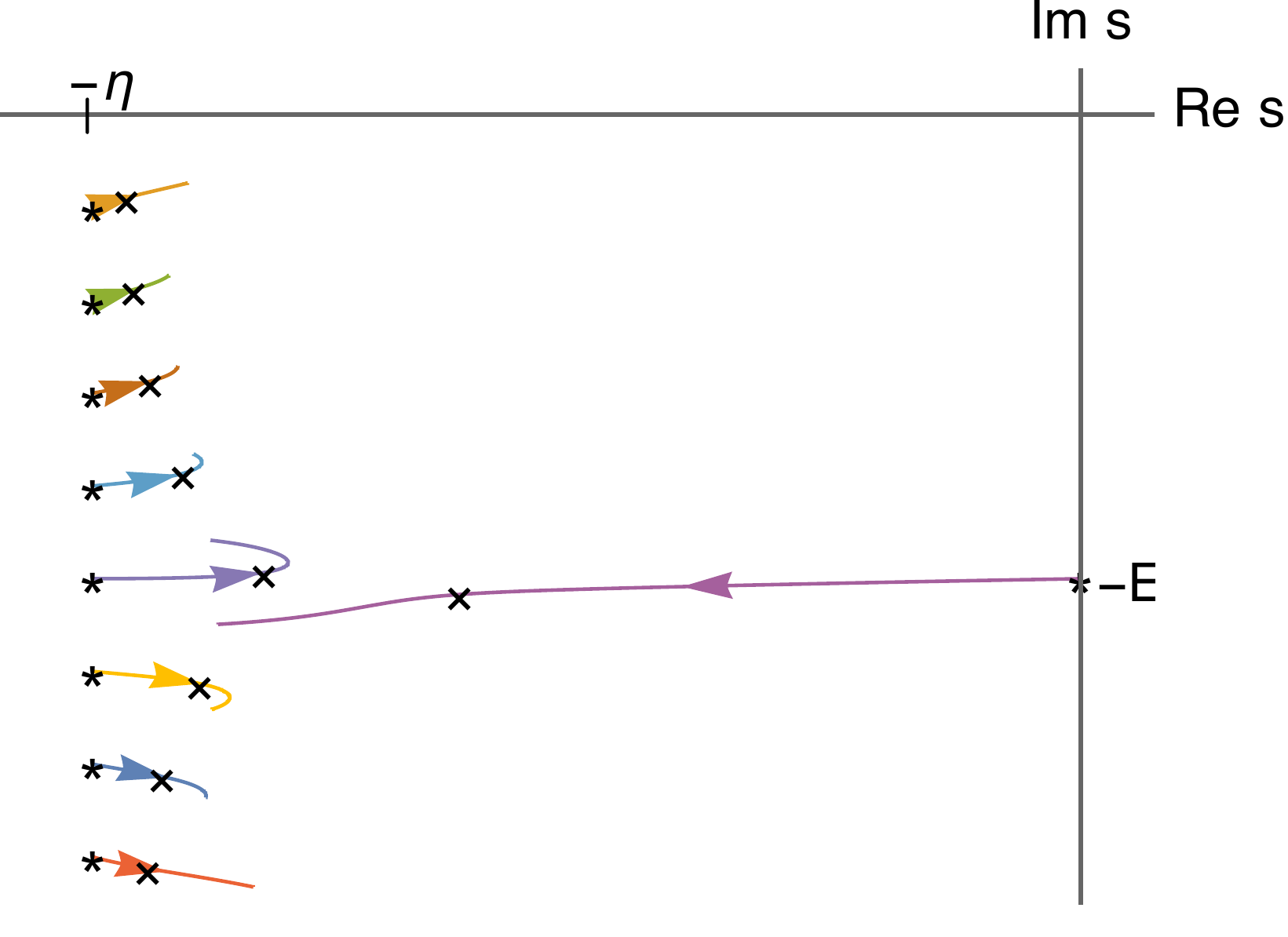}
\caption{\label{fig:poles} Trajectories of poles of ${\tilde C(s)}$ in the complex s-plane for $N=8$ for varying $\bar g$. Phonon damping $\eta$ is small but finite in this sketch.  For ${\bar g}=0^+$, there is a pole (*) at $s_0=-i E$ (entrance channel pole) and $N$ poles (inelastic channel poles) located along ${\rm Re}\ s_n=-\eta$ for $n=1\dots N$, resulting from the interaction with the $N$ vibrational modes.  For increasing ${\bar g}$, the poles (X) move as indicated. }
\label{poles}
\end{figure}

\section{Transition rate}

The transition rate may be obtained using 
\begin{equation}
R=-2\ {\rm Re}\ s_0\approx 2 {\bar g}^2\ {\rm Im}\ \Sigma(E)
\label{rate}
\end{equation}
where $E=E_c+\sum_p n_p\omega_p$, the initial energy of the system. 
In the absence of phonon damping, there are no true transitions, only the excitation of a resonance \cite{dpc92} that returns to the initial state of the system after a {recurrence} time set by the inverse pole spacing.  

For a large number of modes, the sum resulting from Eq.~\ref{rate} may be replaced by an integral; however, this integral is ambiguous as a singularity lies on the integration path.   The inevitable clamping loss confronting the phonons suggests that the phonon frequencies acquire a small imaginary part $\omega_p\to \omega_p-i \eta$ ($\eta\to 0^+$).  This resolves the ambiguity and gives a transition rate $R$   
\begin{equation}
R\approx 2\pi {\bar g}^2\rho {(n(\Omega_s)+1-\lambda(\Omega_s) \Lambda)^2\over (n(\Omega_s)+1)}\Theta(\omega_D-\Omega_s)
\label{R}
\end{equation}
where $\Lambda=\sum_m \lambda_m(n_m+1)$, $\Omega_s=E_c+E_b+\Delta$, $\omega_D$ is the highest vibrational frequency supported by the membrane, and $\Theta$ is the Heaviside function.  As a simple check of this result, the Rabi regime case $R_0$ is immediately recovered from the general result in Eq.~\ref{R} by taking  $g_{bb}\to 0$, while the IBM regime result follows from $\lambda_p\to \frac{g_{bb}}{\omega_p}$, the solution to Eq.~\ref{lambda} for $\alpha < \alpha_c$.

\begin{figure}
\includegraphics[width=12cm]{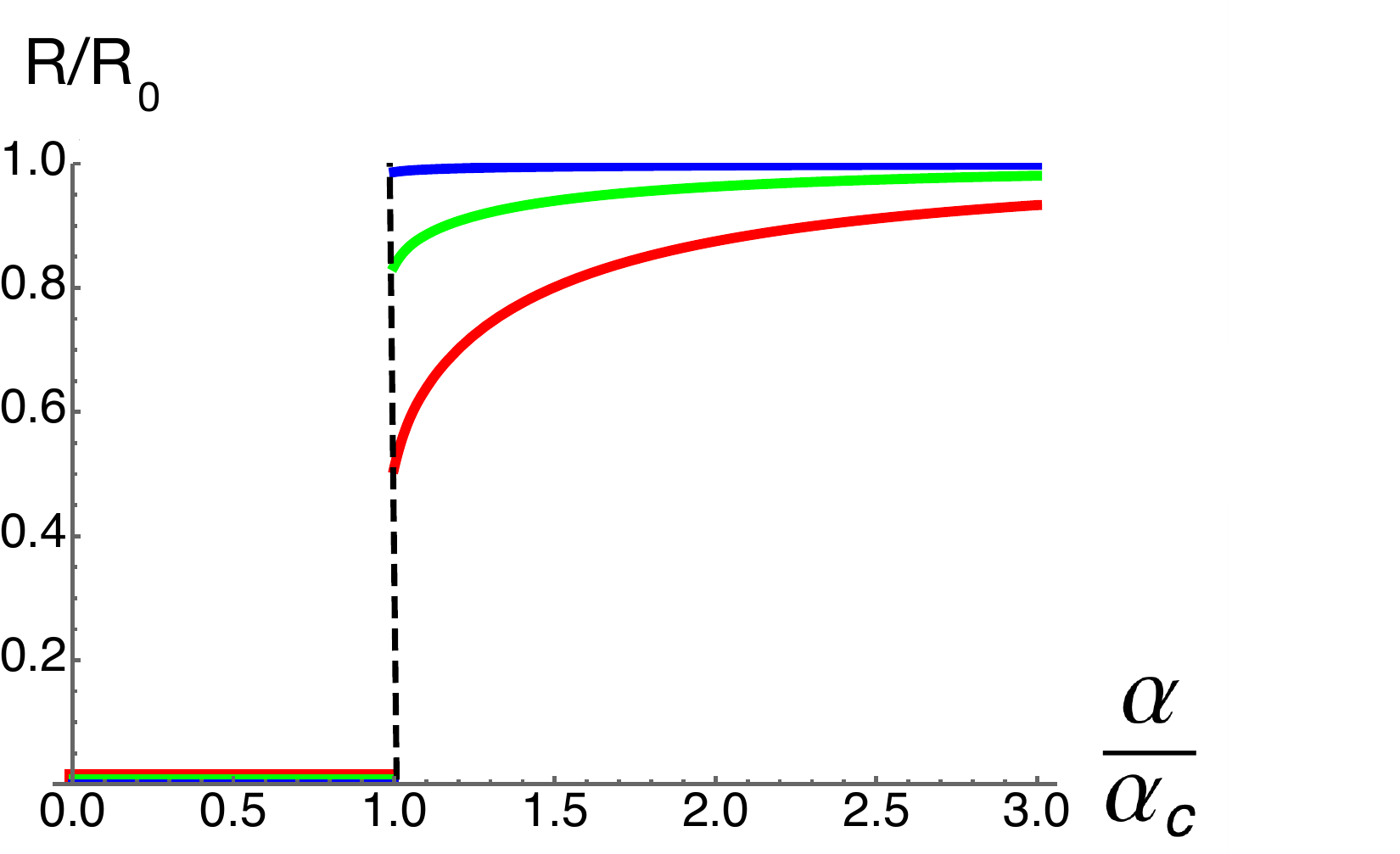}
\caption{\label{fig:gamma} Plot of the  relative adsorption rate  $R/R_0$ versus coupling strength $\alpha/\alpha_c$ for various adsorbent temperatures ($T/g_{bb}= 0.5$ (blue), $2$ (green) and $4$ (red)).  Model parameters used are given in Table~\ref{table:params}. }
\end{figure}

\begin{figure}
\includegraphics[width=12cm]{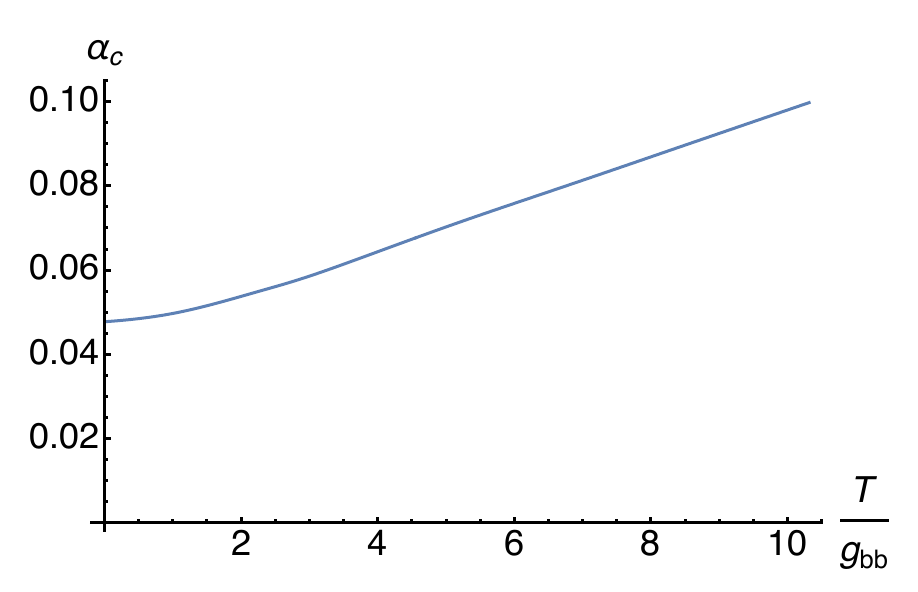}
\caption{\label{fig:alphac} Critical coupling $\alpha_c$ versus temperature $T$ for $N=2000$ modes.}
\end{figure}

The ratio $R/R_0$ is plotted as a function of $\alpha/\alpha_c$ in Fig.~\ref{fig:gamma}.  $\alpha_c$, a slowly varying function of the membrane temperature, is defined as the coupling strength where the self-consistent field  $\Delta_0$ changes discontinuously with $\alpha$ (see Fig.~\ref{fig:alphac}).

\begin{center}
\begin{table}[h]
\caption{\label{table:params} Parameter values for quantum adsorption model.}
\centering
\begin{ruledtabular}
  \begin{tabular}{  c  c  c   c  c  c  c}
    $\alpha$ & $g_{kb}/g_{bb}$ & $\omega_D/g_{bb}$ &$T/g_{bb}$ &$E_b/g_{bb}$ &$\epsilon/g_{bb}$\\ 
\hline
   0.001-0.145  & 0.1-1.2 &  50& 0-10& 10&0.01-0.1\\ 
  \end{tabular}
  \end{ruledtabular}

\end{table}
\end{center}

\section{Feshbach resonances}

$C(t)$ is obtained by inverse transforming the result of Eq.~\ref{C(s)}, giving 
\begin{equation}
C(t)= \sum_{n=0}^{N} A_n e^{s_n t} 
\label{entire}
\end{equation}
where 
\begin{equation}
A_n= \bigg(1+{\bar g}^2 \sum_{p=1}^N {(n_p+1-\Lambda\lambda_p)^2\over (n_p+1)(i s_n+E_b+\Delta-\omega_p-\sum_m n_m\omega_m)^2}\bigg)^{-1}
\end{equation}
{Eq.~\ref{entire} implies a dynamical description of adsorption based on the excitation and decay of a resonance consisting of a coherent, weighted superposition of  contributions from the poles of $\tilde C(s)$, each a Feshbach resonance involving a different vibrational mode (see Fig.~\ref{poles}).}
$A_n$, a normalized distribution in mode number $n$, is localized about the mode whose phonon frequency is approximately $\Omega_s$.   $A_n$ quantifies the contribution of the nth pole to the resonance.  ($A_0$ is the amplitude of the pole associated with the entrance channel).  

\begin{figure}
\includegraphics[width=15cm]{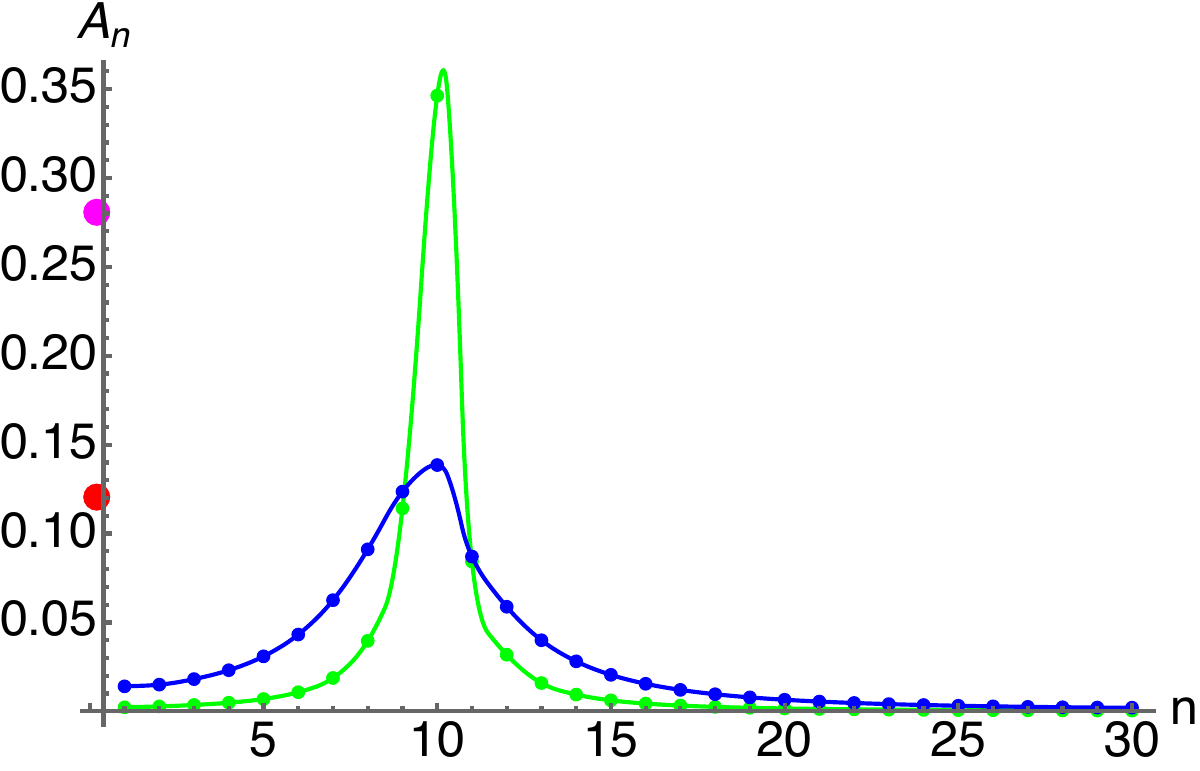}
\caption{\label{fig:An} Amplitude $A_n$ versus mode number $n$ for the {``compound adsorbent''} resonance. Parameter values chosen include $E_b=10$, $E_c=0$, $T=0$, $N=30$, ${\bar g}=0.8$ (blue), and ${\bar g}=0.4$ (green).  The elastic amplitude $A_0$ is plotted as a point (${\bar g}=0.8$ (red) and ${\bar g}=0.4$ (magneta)).}
\label{An}
\end{figure}

The width of the frequency distribution is the resonance decay rate (see Fig.~\ref{An}).  This can be seen by replacing the modal distribution $A_n$ by a Lorentzian of width $\Gamma$ in the continuum limit of Eq.~\ref{entire} and integrating.  This yields $C(t)\sim e^{-i E t} e^{-\Gamma t/2}$, as expected for a decaying resonance.  To complete the adsorption process, the decay of the resonance to the bound state results in the creation of a phonon, a decay product of the unstable resonance. {The mode of the phonon can vary over the width of the resonance with probability $A_n$ for the nth mode.}

{Bohr \cite{bohr} described low-energy nucleon capture as a two-step process: the first step involves the formation of the compound nucleus, an unstable many-body state involving the nucleus and the nucleon, and the second step is the decay of the compound nucleus into reaction products.  In analogy, the adsorption resonance state identified here  that consists of a weighted, superposition of Feshbach resonances plays the role of Bohr's compound nucleus in nucleon capture.  Thus it seems appropriate to dub this analogous state in adsorption the ``compound adsorbent'' resonance.}

{The exponential suppression of the adsorption rate can be understood in terms of the compound adsorbent model.  For $\alpha$ crossing below $\alpha_c$,  the discontinuous change of the self-consistent field $\Delta_0$ leads to an exponential reduction in the effective transitional coupling constant $\bar g$.  The width of the compound adsorbent resonance is then exponentially reduced.  When the width of the narrow resonance becomes comparable to the pole spacing, only the elastic pole contributes substantially to the resonance ($A_0\to 1$), and the inelastic outcomes corresponding to adsorption are no longer possible.}

In summary, 
{the time-dependent variational principle of Dirac and Frenkel can be used to analyze the dynamics of many-body quantum systems nonperturbatively.  In the case of the quantum adsorption model, the equations of motion were found to be a system of first-order ordinary differential equations that were analytically tractable.}
The results of this approach show that the region of intermediate coupling contains a discontinuous change in the adsorption rate, a prediction that could be directly tested experimentally following previously used techniques for measurements on adsorption of hydrogen on superfluid helium \cite{yu93}.  The magnitude of this discontinuity is reduced with increasing temperature.    At low temperature and above a critical coupling $\alpha_c$, the variational results are well approximated by Fermi's golden rule; below the critical coupling, the results show the anticipated {phonon orthogonality catastrophe} effect that suppresses the transition rate by roughly 10 orders of magnitude for the parameter values considered.  

{The time-dependent solution obtained contains a mathematical description of the dynamics that is analogous with Bohr's compound nucleus model for nucleon capture: a resonant state is formed involving the adsorbate and vibrational modes of the adsorbent.  With the decay of the resonant state, there are a variety of possible outcomes resulting in adsorption of the atom with the creation of an accompanying phonon in modes contributing to the resonant state.}  Support of this work under NASA grant number 80NSSC19M0143 is gratefully acknowledged.  


\bibliographystyle{apsrev4-2}
\bibliography{qs,06fred}

\end{document}